\documentclass[aip,preprint]{revtex4-1}
\usepackage{graphicx}

\draft 

\begin{document}

\title{\textit{Ab initio} simulations of the thermodynamic properties and phase transition of Fermi systems based on fictitious identical particles and physics-informed neural networks}

\author{Yunuo Xiong}
\email{xiongyunuo@hbpu.edu.cn}
\affiliation{Center for Fundamental Physics and School of Mathematics and Physics, Hubei Polytechnic University, Huangshi 435003, China}

\author{Hongwei Xiong}
\email{xionghongwei@hbpu.edu.cn}

\affiliation{Center for Fundamental Physics and School of Mathematics and Physics, Hubei Polytechnic University, Huangshi 435003, China}

\date{\today}

\begin{abstract}
Fictitious identical particle thermodynamics has emerged as a powerful tool to overcome the fermion sign problem, enabling highly accurate simulations of one thousand fermions in warm dense matter (T. Dornheim et al., J. Phys. Chem. Lett. 15, 1305 (2024)). However, inferring the thermodynamic properties of Fermi systems from a large number of exact numerical simulations of the bosonic sector still poses subtle challenges, especially in the regime of high quantum degeneracy and in the presence of phase transitions. In this work, we demonstrate that physics-informed neural networks (PINNs), trained on data from extensive and sign-problem-free numerical simulations of the bosonic sector, offer a valuable means to infer the thermodynamic properties of Fermi systems. PINNs can play a particularly crucial role in capturing phase transitions. To illustrate the methodology of fictitious identical particles combined with PINNs for simulating the thermodynamics of Fermi systems, we explore its application in realistic scenarios, including ultracold Fermi gases in periodic potentials, and phase transitions of pair condensation formed in the unitary limit in a three-dimensional harmonic trap. For the spatially continuous Fermi-Hubbard model, we efficiently and reliably simulated hundreds of fermions here. For the Fermi gas in the unitary limit, based on the fictitious identical particle combined with PINNs, our approach confirms  the universal result of the critical temperature with the increasing of the number of fermions, and is consistent with the experimental observations.
\end{abstract}

\maketitle

\section{Introduction}

For general Fermi systems, it is a highly challenging fundamental scientific problem to accurately simulate the thermodynamic properties of Fermi systems, especially phase transitions, from first principles due to the obstacle of the fermion sign problem \cite{ceperley,troyer,Dornheim-PRE,Alex,WDM}.
In a nutshell, when using methods such as path integral Monte Carlo (PIMC) or path integral molecular dynamics (PIMD) to perform \textit{ab initio} numerical simulations of Fermi systems, for a physical quantity $p$, after a large number of samplings, we have
\begin{equation}
p(T)=\frac{A(T)}{s(T)}.
\end{equation}
Here, $s$ is called the sign factor caused by the antisymmetry of the exchange of identical fermions. In general, $s\sim e^{-N/T}$. Here, $N$ refers to the number of fermions and $T$ represents the temperature. With the increase of the number of fermions and the decrease of temperature, the sign factor shows an exponential decay behavior; thus, it leads to a fundamental obstacle to accurately simulating $s$ by sampling for large systems or extremely low temperatures. This fundamental problem in physics is called the fermion sign problem \cite{WDM}.

Over the past few decades, people have developed various methods \cite{nodes,Helium,Militzer,Mak,Rubtsov,Blunt,Malone,Schoof1,Schoof2,Schoof3,Yilmaz,PB1,PB2,Groth,Joonho,SWZhang,QinMP,DornheimMod,diMonte,Burov2,diMonte2} to try to overcome the fermion sign problem. Recently, the thermodynamics of fictitious identical particles \cite{XiongFSP,Xiong-xi} has provided an efficient means to overcome the fermion sign problem to simulate the thermodynamic properties of general fermionic systems, and two methods have been developed: isothermal extrapolation \cite{XiongFSP} and constant energy semi-extrapolation \cite{Xiong-xi}. Based on fictitious identical particles and the isothermal extrapolation method, after carrying out accurate numerical simulations  in the bosonic sector using PIMC, Dornheim et al. verified in two groundbreaking papers \cite{Dornheim1,Dornheim2} that fictitious identical particles are a valuable tool for overcoming the fermion sign problem. The work of Dornheim et al. greatly extended the application of fictitious identical particles in the simulation of fermions, efficiently and highly accurately simulating a series of thermodynamic properties \cite{Dornheim1} of fermionic systems, in particular carrying out simulations in warm dense matter with 1,000 fermions \cite{Dornheim2} for the first time that have a key promoting effect on the National Ignition Facility.

In the thermodynamics of fictitious identical particles \cite{XiongFSP,Xiong-xi}, we introduce a parameter $\xi$ to characterize the fictitious identical particles: $\xi=1$ represents bosons, and $\xi=-1$ represents fermions. We can use PIMD \cite{HirshPIMD,HirshImprove,Deuterium,Xiong-spinor,Xiong-Momentum,Xiong-magnetic,Xiong-anyon,Xiong-Green,XiongFSP,Xiong-xi} or PIMC \cite{Dornheim1,Dornheim2,CeperleyBook,Feynman,Tuckerman,barker,Morita,CeperleyRMP,Burov1,Burov1b} to accurately simulate the bosonic sector ($\xi>0$), that is, to numerically simulate fictitious identical particles with different parameters $\xi$ at different temperatures, and then try to continuously extrapolate the thermodynamic properties of the fermionic system from zero temperature to finite temperature. For a fictitious identical particle with parameter $\xi$, for a physical quantity $p$, we have
\begin{equation}
p(\xi,T)=\frac{A(\xi,T)}{s(\xi,T)}.
\end{equation}
When $\xi\geq 0$, there is no fermion sign problem, so we can carry out accurate numerical simulations to obtain a large amount of data about the physical quantity $p$.

\begin{figure}[htbp]
\begin{center}
 \includegraphics[width=0.8\textwidth]{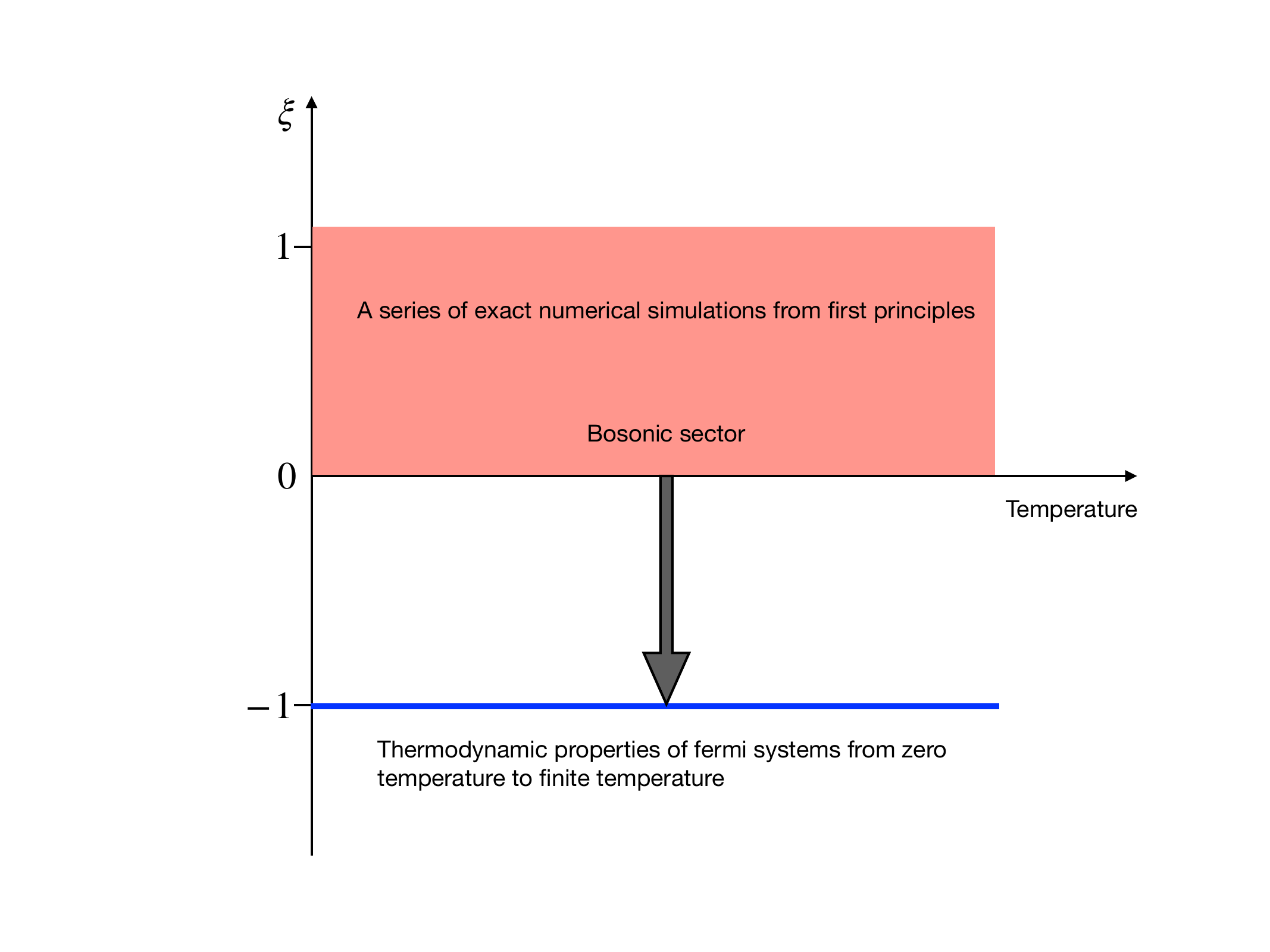} 
\caption{\label{figure1}  
The thermodynamics of fictitious identical particles with a real parameter $\xi$ extends the one-dimensional space of the thermodynamics with respect to temperature to a two-dimensional abstract space that includes both temperature and the real parameter $\xi$. The blue solid line in the figure represents the thermodynamic properties of fermions ($\xi=-1$) at different temperatures. The orange area in the figure represents the region where there is no fermion sign problem and accurate numerical simulations can be performed. This paper will use physics-informed neural networks to extrapolate the thermodynamic properties of fermionic systems.}
\end{center}
\end{figure}

Fig. \ref{figure1} shows the general idea of inferring the thermodynamic properties of fermions based on the thermodynamics of fictitious identical particles. The orange area in the figure represents the bosonic sector where there is no fermion sign problem. First, we obtain a large amount of accurate numerical simulation results from first-principles calculations for different $\xi$ and different temperatures in the bosonic sector ($\xi\geq 0$). Usually, we can use PIMD or PIMC to carry out accurate numerical simulations of fictitious identical particles. After preparing a large amount of these data, we can consider using various methods to infer the thermodynamic properties of the fermionic system. 

In Fig. \ref{figure1}, the blue solid line with $\xi=-1$ represents the thermodynamic properties of the fermionic system. In the paper \cite{XiongFSP}, we proposed the isothermal extrapolation method applicable to medium or weak quantum degeneracy, and in the paper \cite{Xiong-xi}, we proposed the constant energy semi-extrapolation method to overcome the difficulty of inferring the thermodynamic properties of the fermionic system with high quantum degeneracy. For a physical quantity $p(\xi,T)$ related to $(\xi,T)$, based on general considerations of thermodynamic properties, there is a strict relationship \cite{Xiong-xi}  $\left.\frac{\partial p(\xi,T)}{\partial T}\right|_{T=0}=0$. It is this strict relationship that makes the constant-energy semi-extrapolation more accurate and applicable in principle than the isothermal extrapolation.

Despite the existence of two methods, isothermal extrapolation and constant energy semi-extrapolation, it is still necessary to develop other methods to infer the thermodynamic properties of fermionic systems. For some important physical systems, especially when there is a phase transition, there may be complex transformations in the transition from the bosonic sector to the fermionic sector. For example, at zero temperature and without interaction, the function between energy $E(\xi,T=0)$ and $\xi$ is not an analytic function. The constant energy semi-extrapolation partially overcomes this difficulty by bypassing $E(\xi,T)$ in the region of $\xi>-1$ and small $T$. However, if we consider the functional relationship between fermion energy and temperature $E(\xi=-1,T)$, we will notice that this function can be very complex, especially in the case of phase transitions. When there is a phase transition in the fermionic system, if we want to fit the data between the fermion energy $E$ and the temperature $T$ with a polynomial function, we often need to keep at least the high-order terms up to $T^5$. At this time, even if the constant energy semi-extrapolation is expected to give reasonable thermodynamic properties of the fermionic system, we also urgently need to develop other methods besides the constant energy semi-extrapolation and the isothermal extrapolation.

When we try to fit and find the complicated function $E(\xi=-1,T)$ through energy data of the fermionic system at different temperatures, it is easy to face the problem of overfitting; and when we try to accurately extrapolate the complicated function $E(\xi=-1,T)$ from the bosonic sector, we will naturally face similar difficulties. Fortunately, neural networks \cite{neuralbook} provide a way to effectively suppress overfitting and accurately express complex functions. So how to use neural networks to reasonably infer the complex function $E(\xi,T)$? In a neural network, the function $E(\xi,T)$ can be represented by the weights in the neural network. These weights are determined by continuously reducing the loss function through the backpropagation algorithm with a large amount of pre-prepared energy data.

In the context of inferring the thermodynamic properties of fermionic systems using fictitious identical particles combined with neural networks, the exact condition  
$\left.\frac{\partial p(\xi,T)}{\partial T}\right|_{T=0}=0$ can be used when training the neural network. Therefore, the neural network we use here should fully consider the constraints imposed by physical principles. For neural networks trained with such physical information-based constraints, people generally refer to them as physics-informed neural networks (PINNs) \cite{PINN1,PINN2,PINN3}. For this reason, the topic of this work is to combine fictitious identical particles and PINNs to simulate the thermodynamic properties of fermionic systems, especially phase transitions. Based on the systematic study here, we believe that this approach provides a valuable tool for \textit{ab initio} simulation of the thermodynamic properties of fermionic systems.

The structure of this paper is as follows. In Sec. \ref{PINN}, we briefly introduce PINNs and the general idea of combining fictitious identical particles with PINNs. In Sec. \ref{Hubbard}, we present an application to the simulation of hundreds of fermions in the Fermi-Hubbard model in continuous space. In Sec. \ref{PINN-more}, we introduce a general idea of imposing more physical information on PINNs and demonstrate it in the toy model of an ideal Fermi gas in a three-dimensional harmonic potential. In Sec. \ref{unitary}, we simulate the unitary Fermi gas with short-range interactions, which is a highly challenging problem, and obtain results that are consistent with experiments and previous theoretical benchmarks, demonstrating the efficiency, reliability, and generality of fictitious identical particles combined with PINNs. In Sec. \ref{summary}, we give a summary and discussion.

\section{The general idea of fictitious identical particles combined with physics-informed neural networks}
\label{PINN}

Neural networks are essentially constructed by training to give a reasonable inference of new situations through a complex function \cite{neuralbook}. The training is completed by using a large amount of pre-prepared data and some known basic properties, so as to minimize the loss function using the backpropagation algorithm. After the training is completed, the complex function given by the neural network has the hope to infer the properties of new situations.

\begin{figure}[htbp]
\begin{center}
 \includegraphics[width=0.8\textwidth]{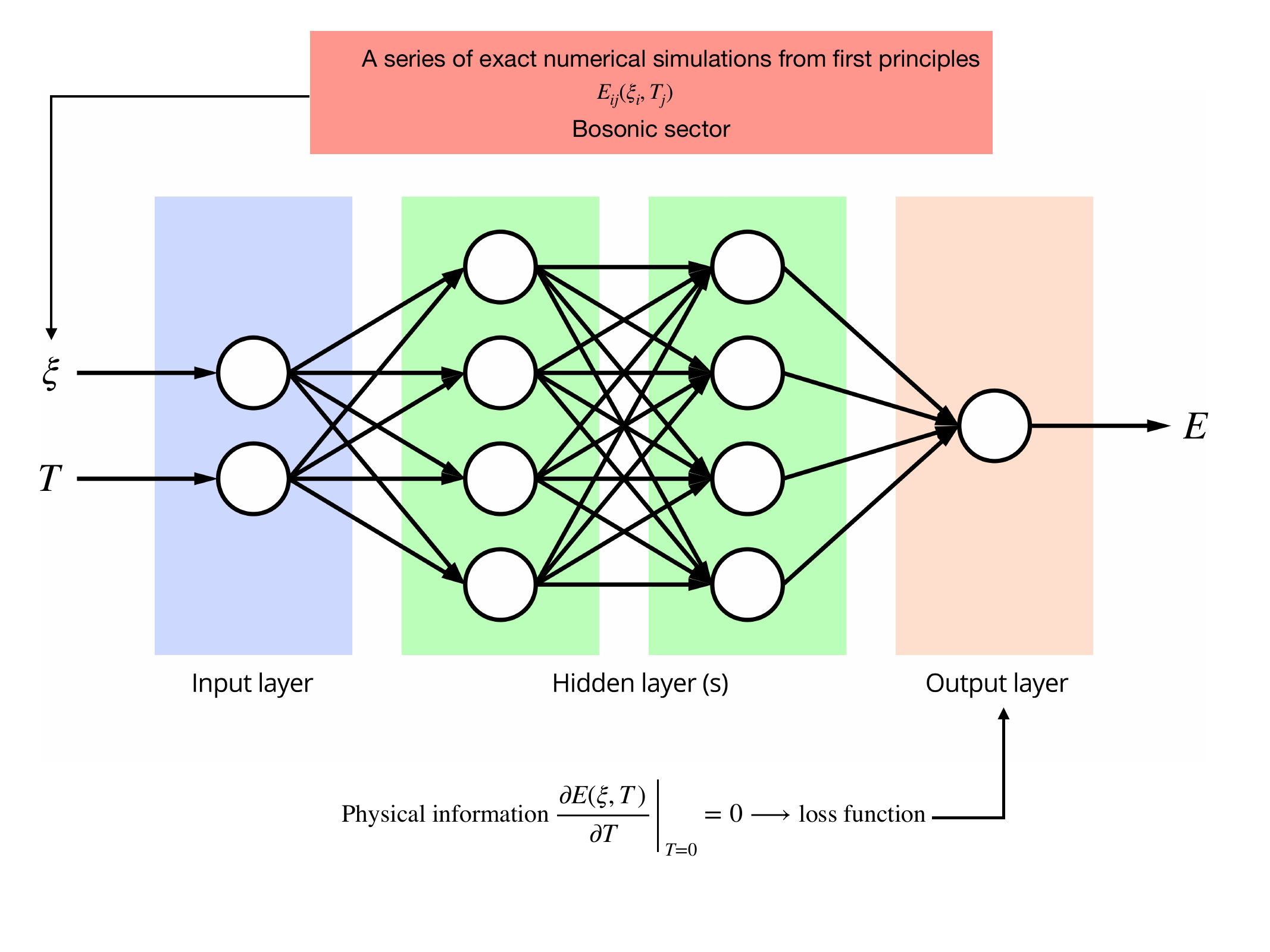} 
\caption{\label{PINN}  
The figure shows the general idea of using physics-informed neural networks to simulate the thermodynamic properties of fermionic systems. In the feedforward neural network, the input is $\xi$ and temperature $T$, and the output is energy $E$. A large amount of data from \textit{ab initio} simulations in the bosonic sector is used as the training data for the neural network.  $(\xi_i,T_j)$ is the input, and the corresponding energy $E_{ij}(\xi_i,T_j)$ is the output, which is the label for neural network training. During the training process, when we add the exact physical information  $\left.\frac{\partial E(\xi,T)}{\partial T}\right|_{T=0}=0$ to the loss function and train it using the backpropagation algorithm, we call this network physics-informed neural network. }
\end{center}
\end{figure}

In the following, we will introduce neural networks briefly based on the thermodynamic inference of fermionic systems using fictitious identical particles \cite{XiongFSP,Xiong-xi}. In Fig. \ref{PINN}, we show a feedforward neural networks (FNNs) that is suitable for the present work. In this neural network, the input is $\xi$ and $T$, and the output is energy $E$. The purpose of training is to find a reasonable energy function $E(\xi,T)$. In fact, before the popularity of neural networks, the physics discipline faced a large number of similar problems. In the traditional methods of physics, we first need to artificially construct a function $E(\xi,T,\theta_k)$, in which there are a series of undetermined parameters $\theta_k$. By comparing with the existing data $\{E_{ij}(\xi_i,T_j),i,j=1,2,3\cdots\}$, we can try to determine these free parameters. The series of undetermined parameters $\theta_k$
can be found by the following extremum condition:
\begin{equation}
L({\theta_k})=\frac{1}{F}\sum_{ij}|E(\xi_i,T_j,{\theta_k})-E_{ij}(\xi_i,T_j)|^2\rightarrow \text{minimum}.
\end{equation}
Here, $F$ refers to the number of data points. We call $L(\theta_k)$ the loss function. The above condition can be partially reflected by the following formula:
\begin{equation}
\frac{\partial L({\theta_k})}{\partial {\theta_k}}=0.
\label{partial}
\end{equation}

However, for the case of large amounts of data and $E(\xi,T)$ being a very complex function, the traditional methods above have several fundamental difficulties: (i) It is difficult for us to artificially construct a reasonable function $E(\xi,T)$. If we artificially expand it using a polynomial function, it is easy to overfit and the extrapolation will completely fail; (ii) For large amounts of data, it is not easy to implement formula (\ref{partial}). (iii) In our research here, there is an additional difficulty with the above idea, that is, the exact condition \cite{Xiong-xi}  $\left.\frac{\partial E(\xi,T)}{\partial T}\right |_{T=0}=0$ is not easy to use in finding the optimal parameters $\theta_k$.

All of the above problems can be well solved in physics-informed neural networks (PINNs) \cite{PINN1,PINN2,PINN3}. In the neural network shown in Fig. \ref{PINN}, the parameters $\theta_k$, which are the weights in the neural network, can be automatically set, as can the loss function. In neural networks, people have also developed the backpropagation algorithm to efficiently and reliably approximate the minimum value of the loss function, and in the training of neural networks, mature methods can be used to well suppress overfitting. In the FNNs shown in Fig. \ref{PINN}, the input is $\xi$ and temperature $T$, and the output is energy $E$. A large amount of data from \textit{ab initio} simulations in the bosonic sector is used as the training data for the neural network. $(\xi_i,T_j)$ is the input, and the corresponding energy $E_{ij}(\xi_i,T_j)$ is the label for neural network training.

After the neural network is trained, the goal is to extrapolate the thermodynamic properties of the fermionic system ($\xi=-1$) from the pre-prepared data in the bosonic sector. However, this extrapolation faces a huge difficulty, that is, there is no information input in the case of $\xi=-1$, which may lead to the failure of extrapolation. Fortunately, based on general considerations of thermodynamic properties, we found an exact relationship:
\begin{equation}
\left.\frac{\partial E(\xi,T)}{\partial T}\right|_{T=0}=0.
\label{exactC}
\end{equation}
This relationship holds for any $\xi$, including negative $\xi$. In the paper \cite{Xiong-xi}, the constant energy semi-extrapolation method based on this exact relationship has been shown to be promising for giving accurate energy predictions from zero temperature to high temperature for some physical systems. In neural networks, it is natural to incorporate this exact relationship as physical information.

So how can the exact relationship (\ref{exactC}) be reflected in the neural network? We just need to add a summation term about this condition to the loss function. Considering this condition, the new loss function is:
\begin{equation}
L({\theta_k})=\frac{\alpha_1}{F}\sum_{ij}|E(\xi_i,T_j,{\theta_k})-E_{ij}(\xi_i,T_j)|^2+\frac{\alpha_2}{Z}\sum_{j=1}^Z\left|\frac{\partial E(\xi_j,T=0,\theta_k)}{\partial T}-0\right|^2.
\end{equation}
Here $\alpha_1$ and $\alpha_2$ are two hyperparameters. In this work, we set $\alpha_1=\alpha_2=1$.

Of course, for the second term on the right side of the above loss function, we need to discretize the partial derivative with respect to $T$. For this term, the summation over $\xi_j$  is performed for different $\xi$ values at $T=0$. In this work, we use 30 different $\xi$ values in the range $-2\leq\xi\leq 1$.

Due to the rapid development of neural networks, people have established mature neural network infrastructure to facilitate the writing of code to implement the above ideas. In this work, we use Python to implement PINNs based on PyToch. In addition, the C code for simulating thermodynamic properties in the bosonic sector using PIMD and the Python code for PINNs are both open source. In this paper, the FNNs we actually used is $\{2,50,50,50,50,50,1\}$. Here 2 represents two inputs and 1 represents one output. 50 represents that the number of neurons in each hidden layer is 50. This simple neural network can be trained quickly.

In the following, we will demonstrate the methods and practical applications of this section for three typical physical systems having practical application value: (1) ultracold Fermi gas in a two-dimensional periodic potential; (2) ideal Fermi gas in a three-dimensional harmonic potential; (3) ultracold Fermi gas in the unitary limit in a three-dimensional harmonic potential.

\section{Ultracold Fermi gases in a two-dimensional periodic potential}
\label{Hubbard}

First, we consider the numerical simulation of ultracold Fermi gas in a periodic potential \cite{lattice,Bloch,Fermilattice}. For fermions in a periodic potential and the simplified Fermi-Hubbard model, the \textit{ab initio} simulation from zero temperature to finite temperature is a very challenging problem \cite{LeBlanc,Qin} due to the difficulty of the fermion sign problem. Here we will find that fictitious identical particle thermodynamics provides a good opportunity to calculate the thermodynamic properties of such a Fermi system from first principles. The joint use of PINNs and the previously developed constant energy semi-extrapolation and isothermal extrapolation can make the value of fictitious identical particle thermodynamics more clearly seen, and confirm whether the simulation results of the thermodynamic properties of the Fermi system are reasonable.

\begin{figure}[htbp]
\begin{center}
\includegraphics[scale=0.5]{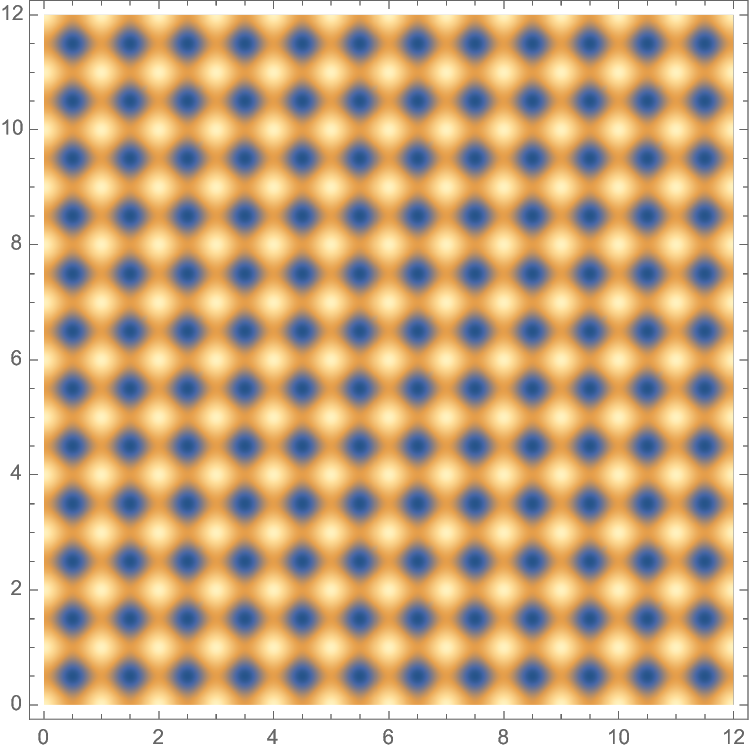}
\caption{
Shown is the two-dimensional periodic potential considered in this paper. More specifically, we also employ periodic boundary condition to study the lattice system with size $L\times L$. In the schematic diagram, the size of the lattice system is $12\times 12$.}
\label{lattices}
\end{center}
\end{figure}

We consider $N_\uparrow$ and $N_\downarrow$  fermions in different spin states, respectively. The total number of fermions is $N=N_\uparrow+N_\downarrow$. We consider the fermions in the following two-dimensional periodic potential:
\begin{equation}
V(x,y)=h\left(\cos^2(\pi x)+\cos^2(\pi y)\right).
\end{equation}
We choose periodic boundary condition to carry out numerical simulations. The size of the box that satisfies the periodic boundary condition is $L\times L$ ($L$ is a natural number greater than zero), and the number of grids contained in the box is $L^2$ . In Fig. \ref{lattices}, we show the $12\times 12$ lattice system that will be studied here.

For $N$ fermions, the Hamiltonian operator is:
\begin{equation}
\hat H=-\frac{1}{2}\sum_{l=1}^N\nabla_l^2+h\sum_{l=1}^N\left(\cos^2(\pi x_l)+\cos^2(\pi y_l)\right)+\sum_{l=1}^{N_\uparrow}\sum_{m=1}^{N_\downarrow}\frac{g}{\pi s^2}e^{-\frac{(\mathbf{r}^{\uparrow}_l-\mathbf{r}^{\downarrow}_m)^2}{s^2}}.
\label{periodic}
\end{equation}
Here $\mathbf{r}^{\uparrow}_l$ and $\mathbf{r}^{\downarrow}_m$ represent the coordinate of the fermion with spin $|\uparrow>$ and  $|\downarrow>$, respectively. We adopt the usual convention $\hbar=k_B=m=1$ in this paper. In the numerical simulation of this paper, the parameters we choose are $h=1, g=1, s=0.5$. 

Even for the simplified two-dimensional Fermi-Hubbard model on a lattice system, due to the limitations of the fermion sign problem, the simulation of hundreds of fermions from zero temperature to finite temperature is still a problem that has not been well solved \cite{LeBlanc,Qin}. For the spatially continuous model considered here, the various methods previously used to simulate the Fermi-Hubbard model will face even greater difficulties. Here we will study a quantum system with $N_\uparrow=N_\downarrow=54$ in a $12\times 12$ lattice system. For such a large-scale case, due to the enormous difficulty of the fermion sign problem, there is a lack of reference for theoretical simulation. Therefore, we need to develop various methods to independently simulate and compare them to judge whether the simulation results are reliable.
 
\begin{figure}[htbp]
\begin{center}
 \includegraphics[width=0.5\textwidth]{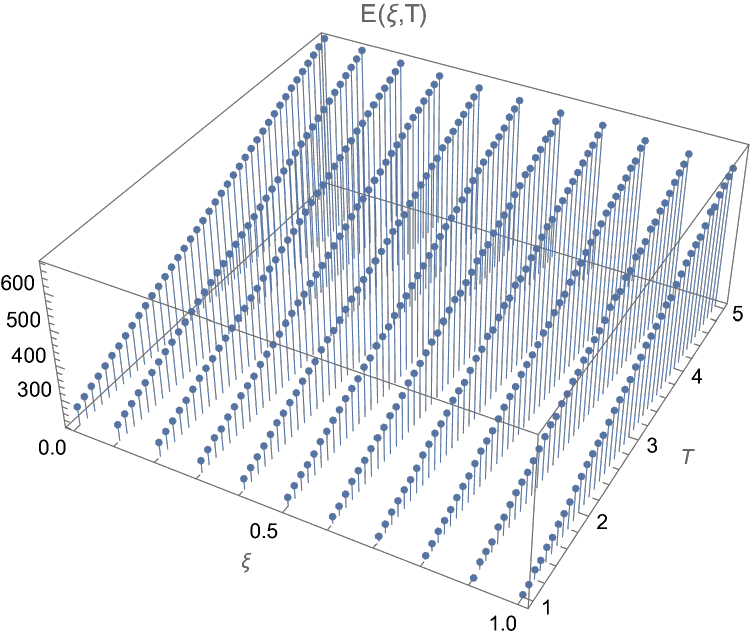} 
\caption{\label{5454data}  The figure shows the energy data of the fictitious identical particles in the bosonic sector obtained by PIMD simulation for a $12\times 12$ lattice with $N_\uparrow=N_\downarrow=54$ in a two-dimensional periodic potential.}
\end{center}
\end{figure}

In Fig. \ref{5454data}, we show the energy data obtained from the exact numerical simulation carried out by PIMD in the bosonic sector. When carrying out the PIMD simulation, we used the recursive method for fictitious identical particles proposed in Ref. \cite{Xiong-quadratic} to generalize the recursive formula of Feldman and Hirshberg \cite{HirshImprove} for the partition function of identical bosons, and used the virial estimator for energy given in our paper.

We will use the energy data shown in Fig. \ref{5454data} as the input to the neural network to train the neural network. When training the neural network, we can judge whether the training is successful by the change of the loss function during the training process and the final value. Since we use very simple FNNs written based on PyTorch, there are not many special skills to declare. The only thing to note is that for the input energy data, we need to normalize it to the range of $[0,1]$ to ensure the success of the training.

\begin{figure}[htbp]
\begin{center}
 \includegraphics[width=1\textwidth]{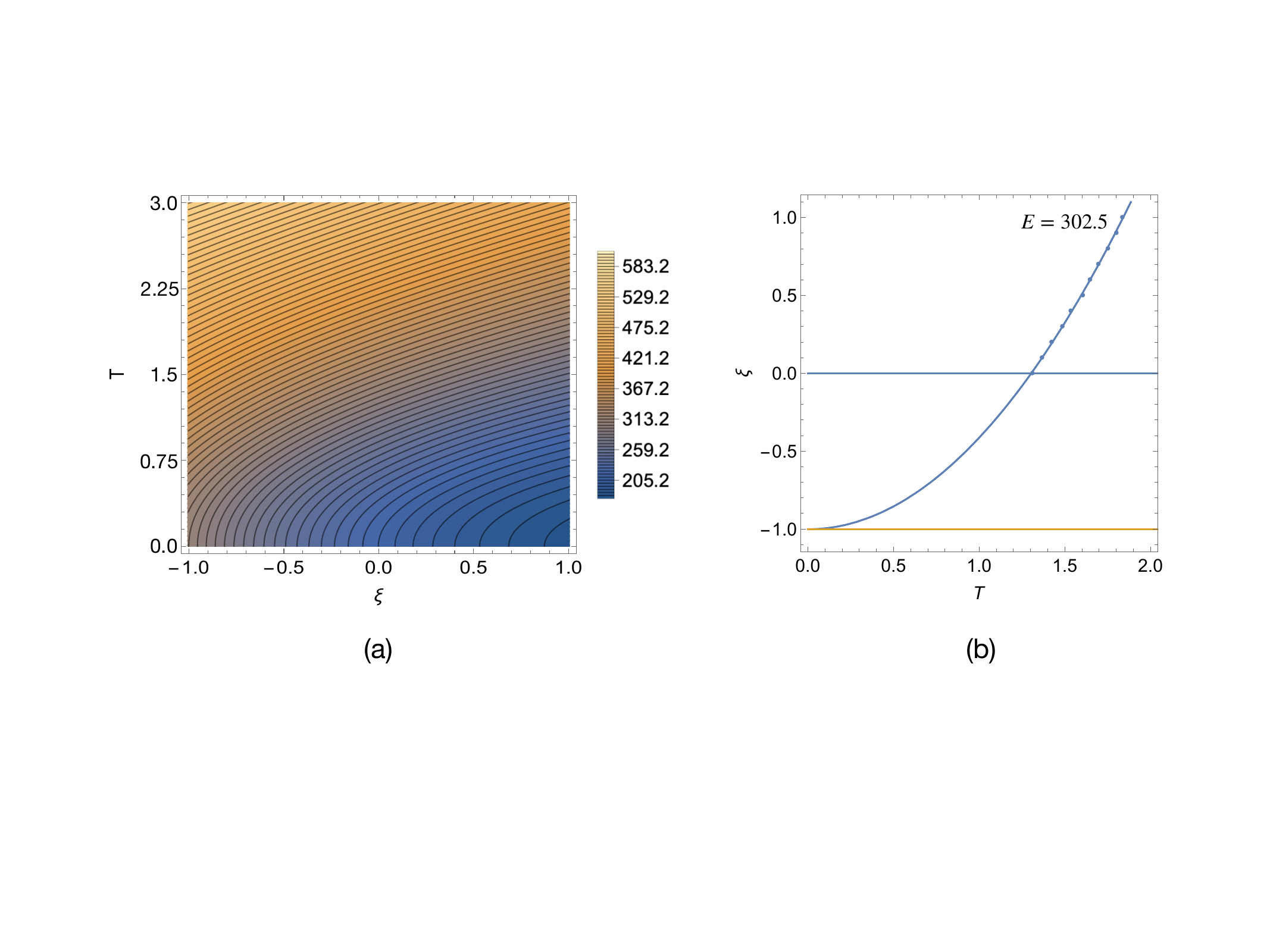} 
\caption{\label{counter5454}  
(a) shows the distribution of the contour lines, which are also the energy curves, of the energy function $E(\xi,T)$ trained by PINNs. (b) shows the energy curve for the ground state energy $E=302.5$ when using the constant energy semi-extrapolation of the fitting function $\xi+d\xi^2=a+b T^2$. }
\end{center}
\end{figure}

After training with PINNs, we can obtain the energy for any $\xi$ at different temperatures. Since the neural network can conveniently give the energy function $E(\xi,T)$ based on the trained weights, in Fig. \ref{counter5454}(a) we give the distribution of the contour lines (energy curves) of the energy function $E(\xi,T)$. We notice that in the high temperature region, these contour lines tend to be straight lines, while in the medium temperature region they are simple parabolas, and in the low temperature region they show the shape of ellipses. In addition, since the physical information  $\left.\frac{\partial E(\xi,T)}{\partial T}\right|_{T=0}=0$ is imposed in the PINNs training, we can clearly see that the contour lines are perpendicular to the $\xi$ axis at $T=0$. In Fig. \ref{counter5454}(a), we notice that the energies in the lower right corner region are almost the same, which is the “black hole” region where the energy is only weakly dependent on temperature and $\xi\geq 0$, as pointed out in Ref. \cite{Xiong-xi}. We find that after PINNs training, we can even give the shape and size of this region, even though we did not prepare the energy data for the bosonic sector when $T<1$ in advance. In the present example, we notice that the ground state energy curve successfully stays away from this “black hole” region, so we can expect that the constant energy semi-extrapolation has the potential to obtain accurate energy results in the whole temperature region for this case. In Fig. \ref{counter5454}(b),  we show the energy curve for the ground state energy $E=302.5$ when using the constant energy semi-extrapolation of the fitting function $\xi+d\xi^2=a+b T^2$. This energy curve coincides with the corresponding contour line in Fig. \ref{counter5454}(a) (the curve that intersects vertically with the point $\xi=-1$ on the $\xi$ axis).

\begin{figure}[htbp]
\begin{center}
 \includegraphics[width=0.7\textwidth]{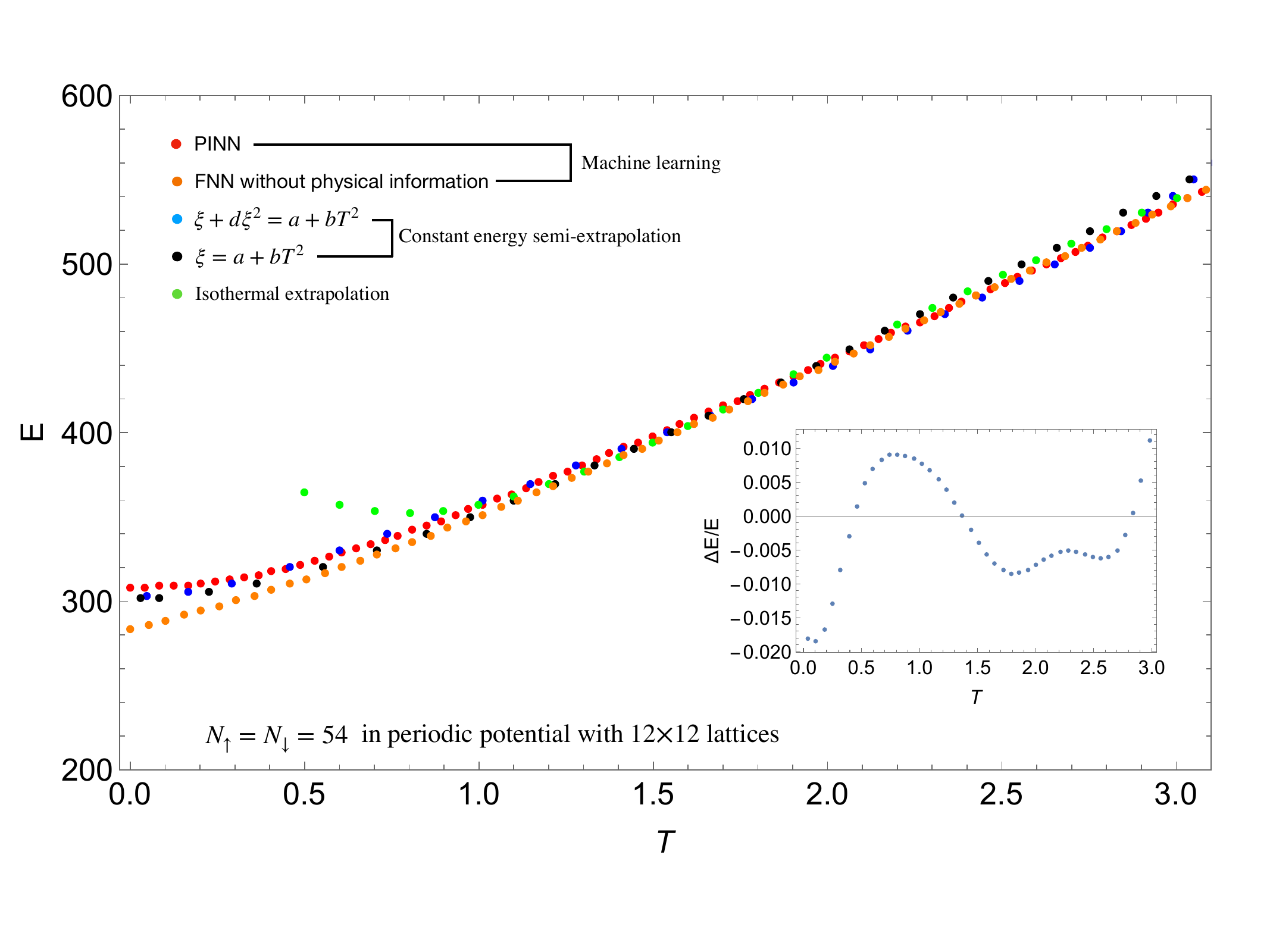} 
\caption{\label{5454energy}  The red dots in the figure represent the results of PINNs, the blue dots are the results of the constant energy semi-extrapolation with the fitting function $\xi+d\xi^2=a+bT^2$, and the black dots are the results of the constant energy semi-extrapolation with the fitting function $\xi=a+bT^2$. The green dots are the results of the isothermal extrapolation. As a comparison with PINNs, the orange dots are the results of directly training the input energy data without imposing the physical information  $\left.\frac{\partial E(\xi,T)}{\partial T}\right|_{T=0}=0$. The inset of the figure shows the relative error of PINNs and the constant energy semi-extrapolation of the fitting function $\xi+d\xi^2=a+bT^2$.}
\end{center}
\end{figure}

The distribution of the contour lines of the energy function given by PINNs inspires us that the constant energy semi-extrapolation has the potential to give accurate energy results in the whole region from zero temperature to high temperature, while the isothermal extrapolation can give accurate energy results in the medium temperature and high temperature regions. In Fig. \ref{5454energy}, we find that this is indeed the case. In Fig. \ref{5454energy}, the red dots represent the results of PINNs, and we notice that the energy given by the red dots is perpendicular to the ordinate when $T\rightarrow 0$, which is in line with physical expectations, because the specific heat should tend to 0 when $T\rightarrow 0$. The blue dots are the results of the constant energy semi-extrapolation with the fitting function $\xi+d\xi^2=a+bT^2$, and the black dots are the results of the constant energy semi-extrapolation with the fitting function $\xi=a+bT^2$. The green dots are the results of the isothermal extrapolation. We notice that when $T>1$, the results obtained by these four methods are relatively consistent. We notice that except for the isothermal extrapolation, the other methods all obtain the correct behavior of the energy of the Fermi system increasing monotonically with temperature.
 
In Fig. \ref{5454energy}, we also show the results of directly training the FNNs without using the  physical information $\left.\frac{\partial E(\xi,T)}{\partial T}\right|_{T=0}=0$  with orange dots. We notice that the results are consistent with those of PINNs at high temperatures. However, at low temperatures, the results of FNNs deviate significantly from those of PINNs, and the specific heat is not 0 when $T\rightarrow 0$. This comparison demonstrates the superiority of PINNs. The reason for this superiority is simple, because the correct physical information is added to the PINNs training. In Fig. \ref{5454energy}, the relationship between energy and temperature given by all different methods shows the monotonic behavior expected by physics. Since all the data in the bosonic sector are adopted in the process of inferring the energy of the Fermi system, the energy fluctuation of the simulation results can be ignored. For this reason, we do not give the error bar caused by energy fluctuation. Of course, there will always be systematic errors in the simulation results compared with the real exact results, for example, the FNNs results have obvious systematic errors at low temperatures. In this example, the consistent results from different reasonable methods can confirm the reliability of the simulation results. By comparing the simulation results of PINNs with the constant energy semi-extrapolation of $\xi+d\xi^2=a+bT^2$, we estimate that the systematic error of the simulation results of PINNs is within 1\%. The inset of Fig. \ref{5454energy} shows the relative error of the PINNs and  the constant energy semi-extrapolation of the fitting function $\xi+d\xi^2=a+bT^2$.

The comparative analysis of the energy results of PINNs, constant energy semi-extrapolation, and isothermal extrapolation once again fully demonstrates the significant application value of fictitious identical particle thermodynamics in important Fermi systems. In principle, based on the powerful  capabilities of supercomputers, we believe that there are no longer fundamental obstacles to the \textit{ab initio} simulations of thousands of fermions in periodic potentials in the future. Since the method of fictitious identical particles combined with PINNs has no special requirements for the spatial dimension of the Fermi-Hubbard model, it is also hoped that the simulation of large-scale three-dimensional Fermi-Hubbard models will become a reality in the near future.

\section{Imposing more physical Information in PINNs and its application to ideal Fermi gas in three-dimensional harmonic potential}
\label{PINN-more}

\begin{figure}[htbp]
\begin{center}
 \includegraphics[width=0.7\textwidth]{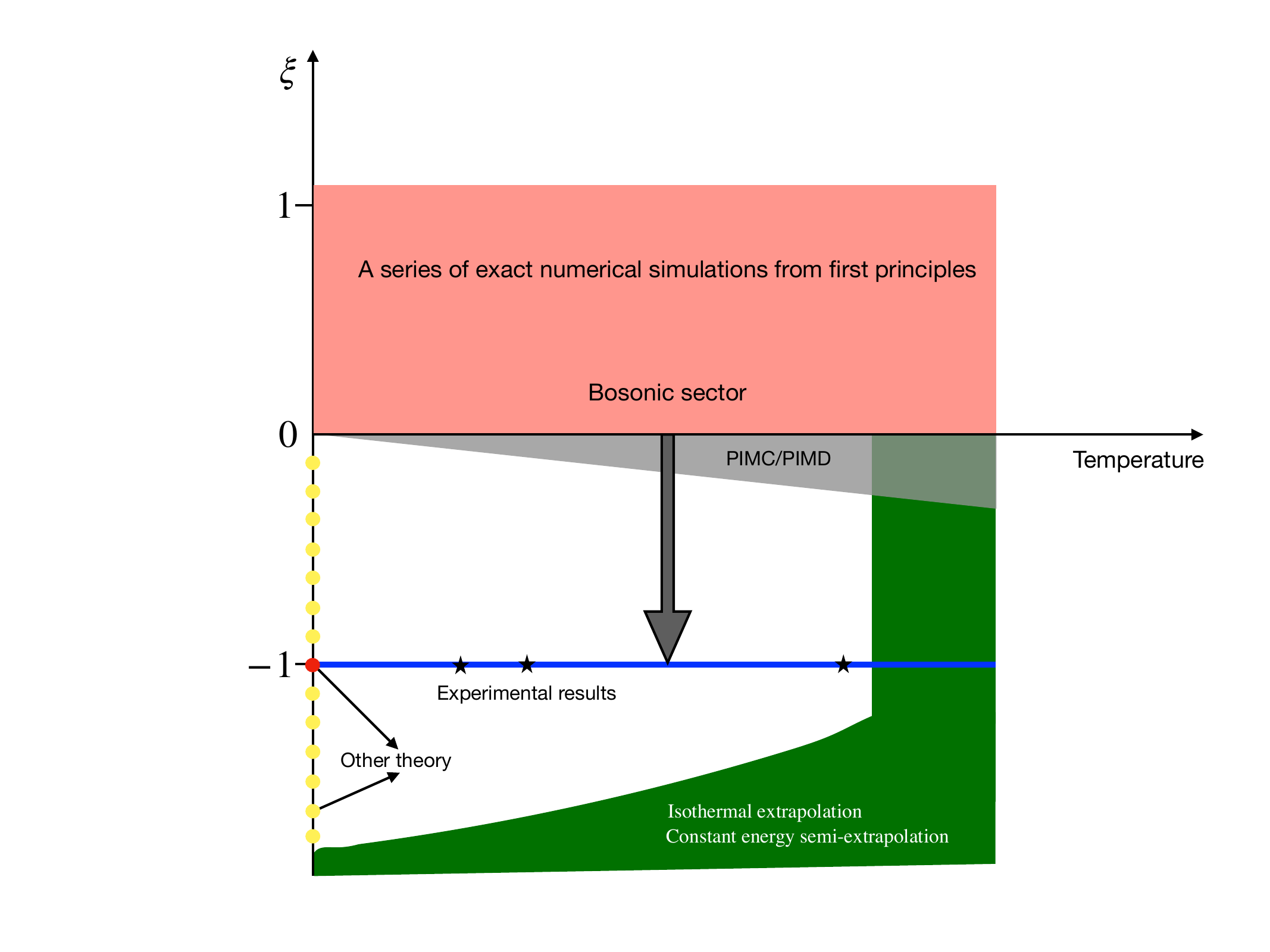} 
\caption{\label{moreinformation}  The orange region in the figure represents a large amount of accurate simulation data for $\xi\geq 0$. The green block represents information from the constant-energy semi-extrapolation and isothermal extrapolation. The gray block represents data that can be directly simulated by methods such as PIMC/PIMD when the fermion sign problem is not so serious. The red dots represent the simulation results of various reliable methods for the ground-state properties of fermions. The yellow dots represent the simulation results of possible future methods for the ground-state properties of general $\xi<0$. The star symbols represent reasonable experimental input data. It is certainly promising to reasonably add some of this information to the input data of PINNs training to improve the accuracy of the simulation of the thermodynamic properties of Fermi systems.}
\end{center}
\end{figure}

In the previous section (Sec. \ref{Hubbard}), we demonstrated the application of PINNs to fermions in periodic potentials, and particularly emphasized the functionality resulting from the incorporation of physical information into PINNs. During the training of the neural network, physical information can also be incorporated into PINNs in various ways. For example, if we know the ground state energy $E(\xi=-1,T=0)$ through theoretical simulation or experimental observation in advance, we can add it to the input data of the neural network for training. In conjunction with Fig. \ref{moreinformation}, the following analyzes the rich physical information that we can impose on PINNs in the most ideal case.

1. For medium and high temperatures, since the nature of the energy contour lines is usually relatively simple, accurate thermodynamic properties of the Fermi system can also be obtained through constant energy semi-extrapolation or isothermal extrapolation. We can use the data of the thermodynamic properties inferred in this way as the input data for PINNs training. We can even extend the entire energy contour line to $(\xi<-1,T=0)$ and include the entire curve in the input data of PINNs. In Fig. \ref{moreinformation}, we use the green block to represent this physical information.

2. For $\xi<-1$ but close to 0, we can directly simulate it using the PIMD/PIMC method, because the fermion sign problem is not so serious at this time. In the papers \cite{XiongFSP,xiong-arxiv} and \cite{Dornheim1}, methods for directly simulating $\xi<-1$ when the fermion sign problem is not particularly serious are given for PIMD and PIMC, respectively. We also found that if the simulation data of $\xi<-1$ is included, the accuracy of the thermodynamic property inference can be improved when using isothermal extrapolation. In Fig. \ref{moreinformation}, we use the gray block to indicate this physical information.

3. Since the establishment of quantum mechanics, people have developed various powerful methods, such as the variational method, to calculate the ground state properties of Fermi systems from first principles. We can consider incorporating the reliable ground state properties obtained by these methods into the input data of PINNs during PINNs training. In Fig. \ref{moreinformation}, we use red dots to represent the ground state properties obtained by these other methods. After determining the ground state energy of the Fermi system, we can even consider incorporating the energy contour line corresponding to the Fermi ground state energy as physical information into the input of the neural network training. In the future, it may also be possible to simulate the ground state properties of general $\xi<0$ using methods such as the variational method, thus providing more information, which we represent by yellow dots in Fig. \ref{moreinformation}. By the way, in the previous application of PINNs, we have already imposed the physical information  $\left.\frac{\partial E(\xi,T)}{\partial T}\right|_{T=0}=0$ on the $\xi$ axis at $T=0$.

4. In experiments, people may also provide some valuable data, which can also be input into the PINNs data in some reasonable way. In Fig. \ref{moreinformation}, we use star symbols to represent.

It can be expected that reasonably adding some of this information to the input data of PINNs training will hopefully improve the accuracy of the simulation of the thermodynamic properties of Fermi systems. In the most ideal case, we note that the horizontal axis of $\xi=-1$ of the thermodynamic properties of the Fermi system of concern is surrounded by various applied physical information, so that it has the characteristics of interpolation during PINNs training. Since this interpolation is performed in a two-dimensional space, and the curve of $E(\xi=-1,T)$ may still be very complex in some cases, this interpolation fitting needs to rely on a powerful tool like PINNs.
 
Below, we analyze the improvements in the simulation accuracy of the Fermi system that can be achieved by imposing various physical information on PINNs for a toy model with an exact solution. We consider non-interacting fictitious identical particles in a three-dimensional harmonic potential. This seemingly simple model can very well reveal the effects of imposing various physical information in PINNs. The reason why this toy model is chosen is also because there is an exact solution that can be compared, so that the degree of systematic deviation caused by the application of PINNs can be grasped.

For a quantum system of $N$ non-interacting two-component particles, the Hamiltonian operator is:
\begin{equation}
\hat H=-\frac{1}{2}\sum_{l=1}^N\nabla_l^2+\frac{1}{2}\sum_{l=1}^N\textbf{x}_l^2.
\label{ideal}
\end{equation}
We adopt the usual convention $\hbar=k_B=m=1$ here. Next we study the case with number of fermions $N_\uparrow=N_\downarrow=10$.

This seemingly simple toy model, however, has a more serious fermion sign problem than the Fermi-Hubbard model. In a periodic potential, the lattice leads to a suppression of the exchange between different fermions. However, in the harmonic potential, the exchange effect between fermions is enhanced due to the external potential that confines all particles in the same small spatial region. This leads to the need for a more complex function to describe the transition of thermodynamic properties between bosons and fermions. For this reason, we expect that unlike the continuous Fermi-Hubbard model analyzed previously, the constant energy semi-extrapolation will have trouble near zero temperature.

\begin{figure}[htbp]
\begin{center}
 \includegraphics[width=0.7\textwidth]{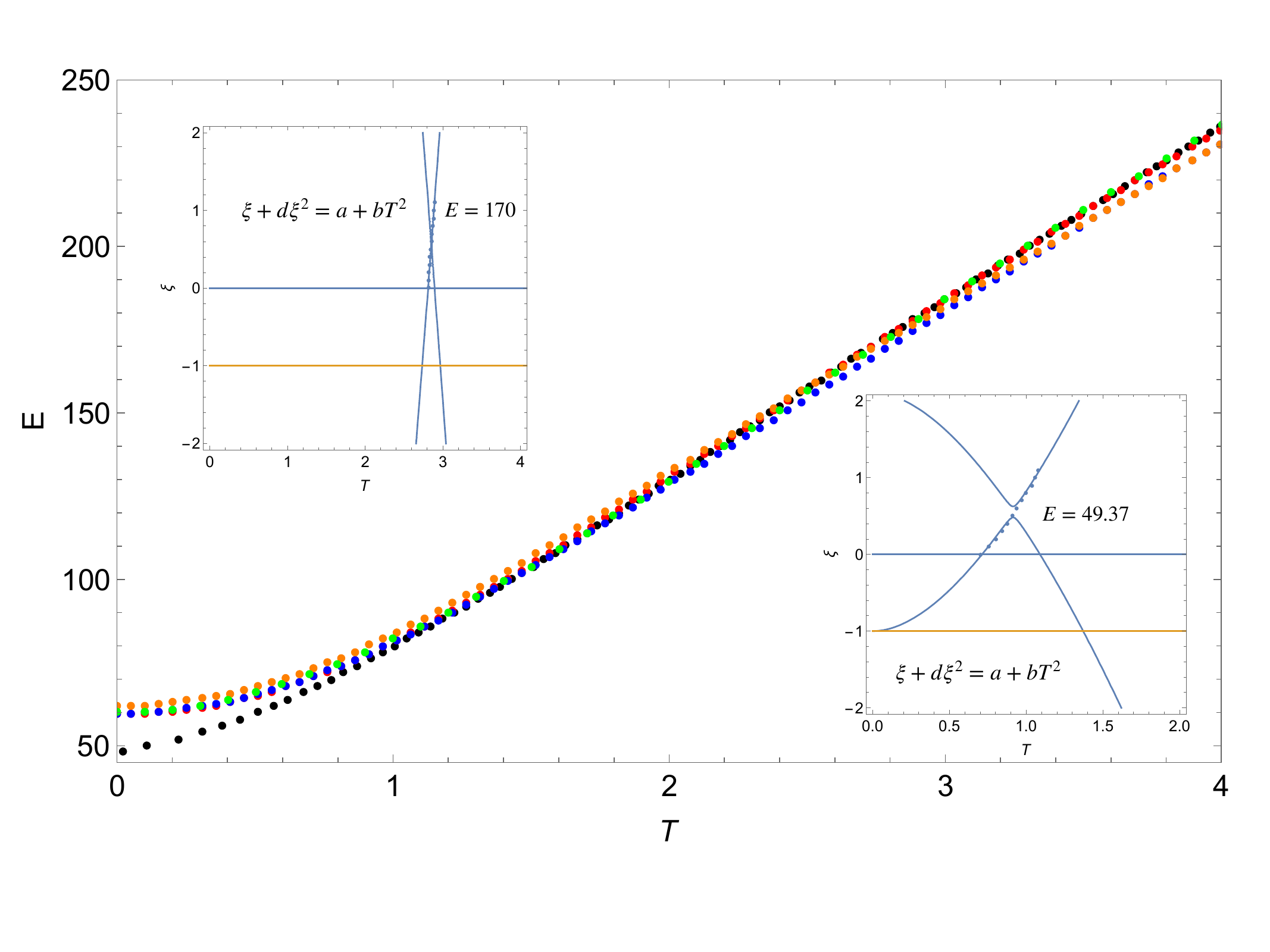} 
\caption{\label{1010idealmore}  The green dots in the figure represent the exact energy, and the black dots represent the results of the constant energy semi-extrapolation using the fitting function $\xi+d\xi^2=a+b T^2$. The ground state energy is $E=60$. In the inset at the lower right corner, we show the constant energy semi-extrapolation curve for $E=49.37$, and in the inset at the upper left corner, we show the fitting situation for medium temperature. The orange dots in the figure are the energy predictions trained by PINNs, the red dots are the energy results after the constant energy semi-extrapolation data of $2<T<4$ is added to PINNs, and the blue dots are the energy results when only the ground state energy data is added during PINNs training.}
\end{center}
\end{figure}

In Fig. \ref{1010idealmore}, the green dots represent the exact energy, and the black dots represent the results of the constant energy semi-extrapolation using the fitting function $\xi+d\xi^2=a+b T^2$. The ground state energy is $E=60$. Due to the extremely serious fermion sign problem, we note that the constant energy semi-extrapolation is clearly unreasonable at temperatures $T<1$, especially at the ground state there is a significant deviation of 18\%, but it is well satisfied at temperatures $T>1$. We also tried other fitting functions to perform the constant energy semi-extrapolation, and there was no significant improvement at low temperatures. In the inset at the lower right corner, we show the constant energy semi-extrapolation curve for $E=49.37$, and we find that there is a clear fitting problem. In the upper left corner of the figure, we show the constant energy fitting situation for $E=170$ at medium temperature, and we find that the fitting function fits the data with $\xi\geq 0$. This explains why the constant energy semi-extrapolation holds at $T>1$ but fails at lower temperatures. This analysis also gives us a standard to judge whether the constant energy semi-extrapolation is effective.

The orange dots in Fig. \ref{1010idealmore} are the energy results obtained by PINNs. We note that there is a slight difference in the ground state energy, and the difference is more obvious at high temperatures. Since the results of the constant energy semi-extrapolation are very accurate at medium and high temperatures, we can consider using the medium and high temperature data of the constant energy semi-extrapolation as physical information for the training of the neural network. The red dots in the figure are the energy results after the constant energy semi-extrapolation data of $2<T<4$ is added to PINNs, which perfectly match the exact results in the whole temperature range, and the average deviation in the whole temperature range is 0.1\%. As a comparison, the blue dots are the results when only the ground state energy data is added during PINNs training, which coincides with the exact energy at $T<2$, but there is a noticeable deviation at $T>2.5$.

It is worth mentioning that the initialization step in the training process of the neural network involves randomness, so the results of each training can be different. This provides a way to get better results through ensemble averaging. This ensemble averaging has already played an important role in the practical application of deep learning. The red dots in Fig. \ref{1010idealmore} are actually the average values obtained after ten trainings.

\begin{figure}[htbp]
\begin{center}
 \includegraphics[width=0.7\textwidth]{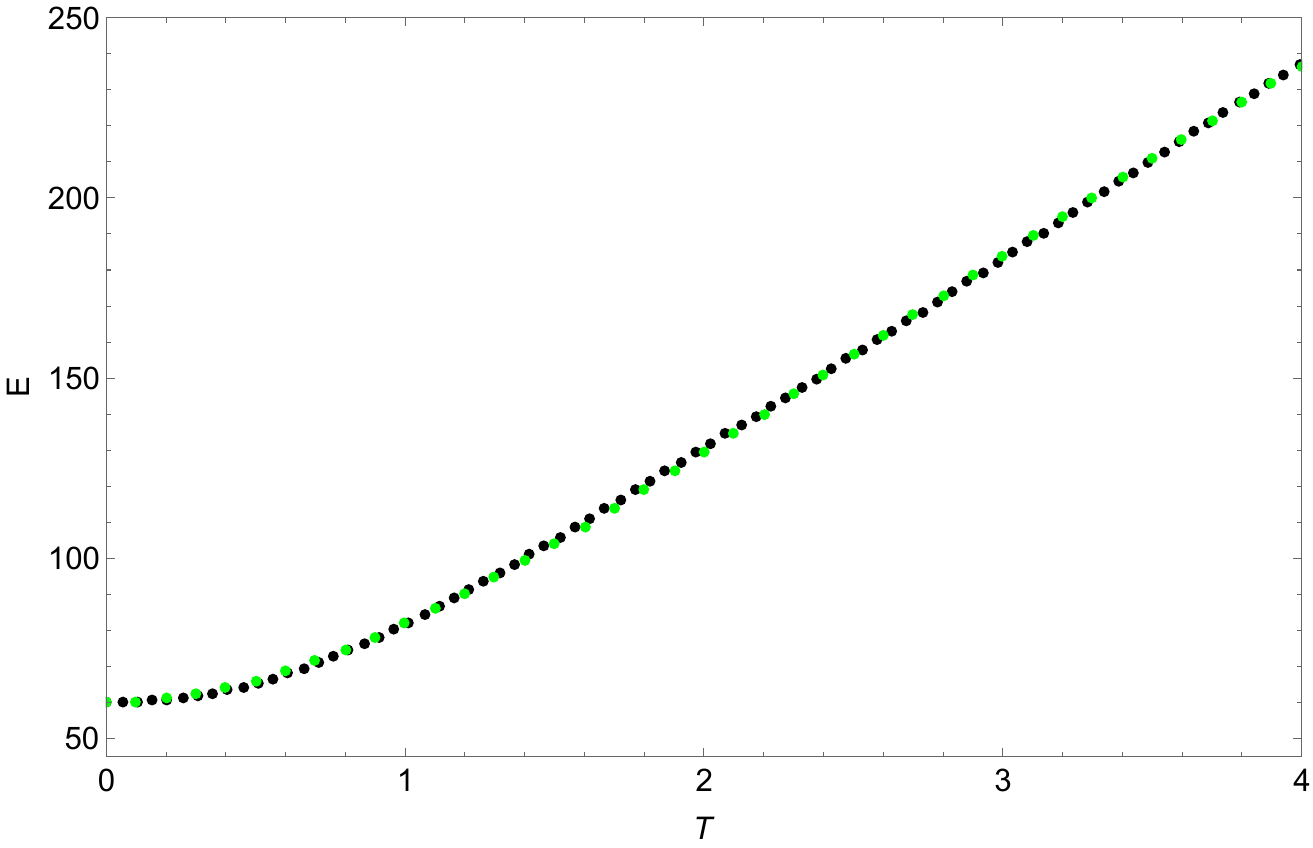} 
\caption{\label{PINNmore}  The green dots in the figure represent the exact energy, and the black dots are the energy results of PINNs training by adding the exact ground state energy and the constant energy semi-extrapolation of $2<T<4$.}
\end{center}
\end{figure}

In Fig. \ref{PINNmore}, we show the energy results of PINNs training with the exact ground state energy and the constant energy semi-extrapolation of $2<T<4$ added simultaneously in the PINNs training. We can clearly see that it fits very well in the whole temperature range, with an average deviation of 0.05\% compared to the exact energy.

In this physical model, we did not observe any phase transitions in our simulations of the fermionic system, which is consistent with physical expectations.
To further demonstrate the value of the combination of fictitious identical particles and PINNs in simulating phase transitions of fermionic systems, we next study the ultracold Fermi gas in the unitary limit in a three-dimensional harmonic potential.

\section{Phase transition simulation of ultracold Fermi gas in the unitary limit in a three-dimensional harmonic potential}
\label{unitary}

For a two-component Fermi gas, the ultracold Fermi gas in the unitary limit with a divergent scattering length for short-range interactions is a strongly correlated quantum system. Since the absolute value of the s-wave scattering length $|a_s|\rightarrow\infty$ is much larger than any other spatial scale, people expect universal behavior and have proposed the Bertsch parameter\cite{Bertsch} to describe the simple and universal relationship between the ground state energy of the Fermi system and the ground state energy of the non-interacting Fermi gas. People have carried out extensive theoretical\cite{diMonte,Ho,Burov2,Strinati} and experimental studies\cite{Thomas,Sagi,Ku,Horikoshi,LiX} on the ultracold Fermi gas in the unitary limit. In particular, in the experimental study of Ku et al.\cite{Ku}, for the Fermi gas in the unitary limit in a three-dimensional harmonic potential, they observed a phase transition leading to a specific heat peak and a universal critical temperature caused by the divergent scattering length.

To clearly demonstrate the value of fictitious identical particles combined with PINNs in simulating phase transitions of fermionic systems, we here specifically analyze the fermionic system in the unitary limit in a three-dimensional harmonic potential. The reason for choosing this quantum system is that although there is still a lack of research on the simulation of finite-temperature thermodynamic properties especially phase transitions in harmonic potentials \cite{Gilbreth}, there are important benchmarks in terms of ground state energy\cite{latticeMC,Mukherjee,Carlson,FNDMC,GFMC}, which can be considered as physical information for PINNs. In addition, experiments\cite{Ku} have also observed phase transitions in three-dimensional harmonic potentials with $10^7$ fermions. Although the number of fermions analyzed in this work is much smaller, the phase transitions observed in the experiment are still valuable for the research in this paper. Moreover, the quantum systems with dozens of fermions analyzed here are expected to be realized in experiments of ultracold Fermi gases in each lattice site of optical lattices.

In a three-dimensional harmonic trap with attractive interaction, the dimensionless Hamiltonian operator is
\begin{equation}
\hat{H}=-\frac{1}{2}\sum_{j=1}^N\Delta_j+\frac{1}{2}\sum_{j=1}^N\textbf{x}_j^2-\sum_{j=1}^{N_\uparrow}\sum_{j'=1}^{N_\downarrow} \frac{ V}{\gamma^2} e^{-|{\textbf x}_j-{\textbf x}_{j'}|^2/\gamma^2}.
\end{equation}
We use the usual convention $\hbar=k_B=m=\omega=1$ here.

The Gaussian interaction in the above Hamiltonian operator can easily adjust the s-wave scattering length $a_s$. To assure the situation of divergent scattering length, the parameter in this paper is chosen as \cite{GaussianInt}
\begin{equation}
\gamma=0.01,~~~~~V=2.684.
\end{equation}
For this choice of $\gamma$, the attractive interaction is short-range and much smaller than the average distance between the fermions studied here, which satisfies the condition of the unitary limit $\gamma<<l<<a_s$. Here $l$ represents the average distance between particles.

Similar to the spatially continuous Fermi-Hubbard model analyzed above, we first accumulate a large amount of energy simulation data in the bosonic sector using PIMD. Then, we can obtain the ground state energy by the constant energy semi-extrapolation method. After these are done, we input the ground state energy (the constant energy semi-extrapolation or the ground state energy in the paper \cite{latticeMC}) and all the energy simulation data in the bosonic sector into PINNs to train and obtain the relationship between the energy and temperature of the Fermi system. In order to study the phase transition, we can judge whether there is a phase transition and obtain the critical temperature by calculating the specific heat according to the derivative of the energy function with respect to temperature.

\begin{figure}[htbp]
\begin{center}
 \includegraphics[width=0.7\textwidth]{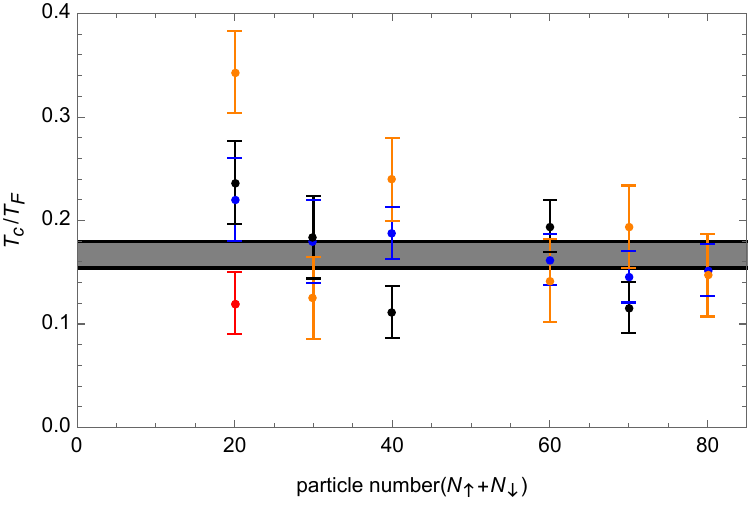} 
\caption{\label{criticaltemperatureN}  When the ground state energy given by the constant energy semi-extrapolation is included in the PINNs training, the blue dots in the figure are the critical temperatures obtained according to the PINNs training results, and the black dots in the figure are the critical temperatures obtained using the ground state energy in the literature \cite{latticeMC} in the PINNs training. The gray area in the figure represents the critical temperature $T_c/T_F=0.167(13)$ observed in the experiment by Ku et al. \cite{Ku}. The red dot in the figure is the critical temperature given for $N=20$ based on contact interaction in the article \cite{Gilbreth}. The orange dots in the figure are the results of extrapolating the critical temperature of the bosonic sector to the critical temperature of the Fermi system.}
\end{center}
\end{figure}

In order to study the phase transition involved in a more systematic way, we studied it separately for different numbers of particles. In Fig. \ref{criticaltemperatureN}, the blue dots are the critical temperatures obtained after PINNs training when the input ground state energy is the constant energy semi-extrapolation result. In the figure, the gray area represents the critical temperature $T_c/T_F=0.167(13)$ observed in the experiment by Ku et al. \cite{Ku}. Here $T_F$ 
is the Fermi temperature of the ideal Fermi gas. We notice that as the number of particles increases, the simulated critical temperature begins to agree with the critical temperature observed in the experiment. In Fig. \ref{criticaltemperatureN}, we also give the critical temperature (black dots) obtained when the ground state energy simulated by lattice Monte Carlo \cite{latticeMC} using contact interaction is input into PINNs. We notice that both slightly different physical information inputs lead to similar results.

\begin{figure}[htbp]
\begin{center}
 \includegraphics[width=0.7\textwidth]{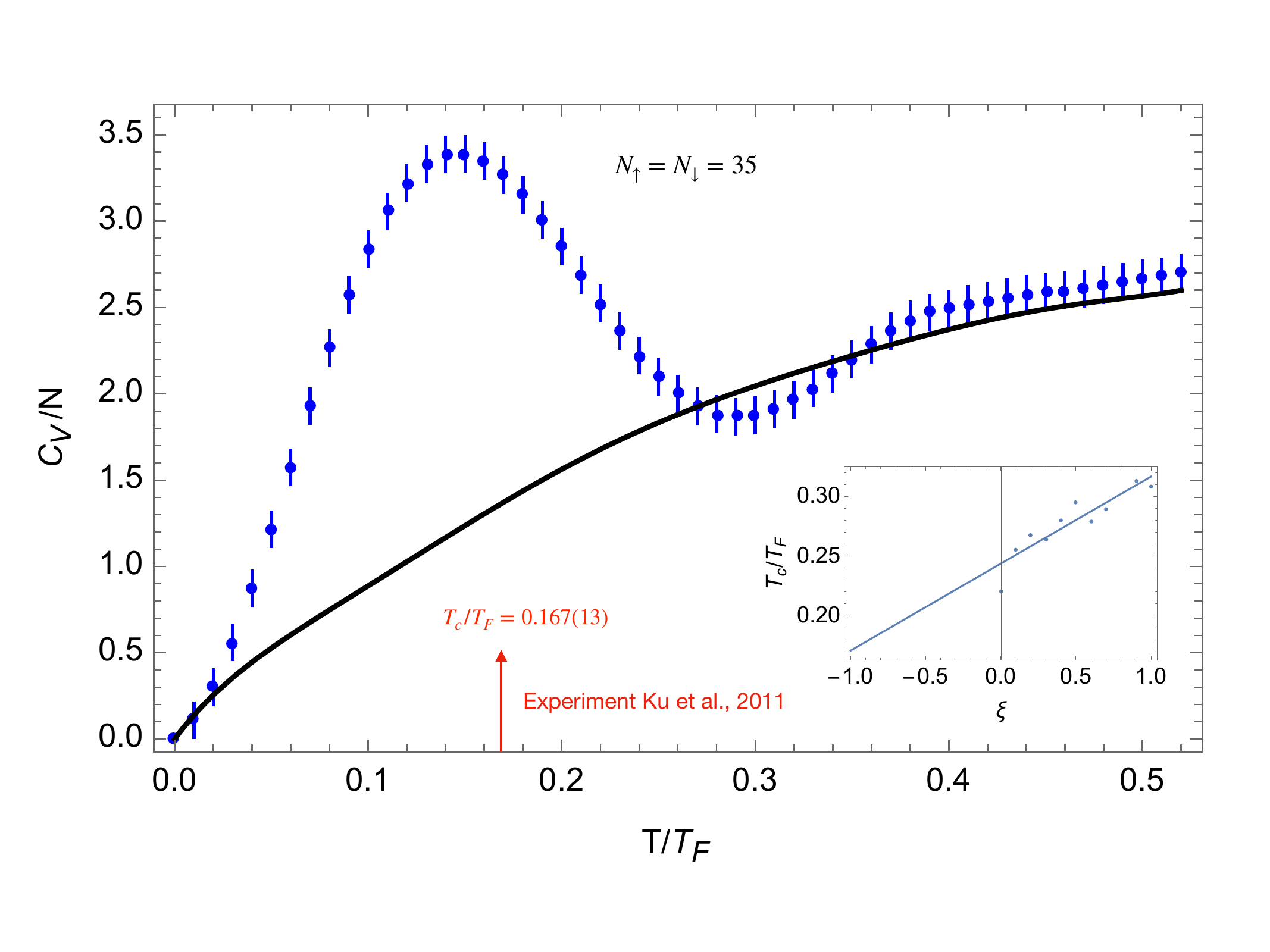} 
\caption{\label{3535specific} For the Fermi system with $N_\uparrow=N_\downarrow=35$ particles, the blue dots in the figure are the specific heat per particle obtained by using PINNs to train the Fermi gas in the unitary limit. As a comparison, the black line in the figure represents the specific heat per particle of the ideal Fermi gas. The red arrow represents the critical temperature observed in the experiment \cite{Ku}. In the inset, we show the process and result of obtaining the critical temperature of the Fermi system through linear fitting of the critical temperature obtained from the bosonic sector.
}
\end{center}
\end{figure}

In Fig. \ref{criticaltemperatureN}, we also give the critical temperature of the Fermi system (orange dots) based on the idea in the article \cite{Xiong-quadratic}. For the bosonic sector, we can obtain the critical temperature of the phase transition for different $\xi$. Thus, we can get a set of data $\{\xi_j, T_c(j)\}$. As suggested in the article \cite{Xiong-quadratic}, we hope to obtain the critical temperature of the Fermi system by extrapolating this set of data about the critical temperature. In the inset of Fig. \ref{3535specific}, we give the extrapolation process and results for the Fermi system with $N_\uparrow=N_\downarrow=35$. This method predicts the critical temperature for the formation of the pair condensation to be $T_c/T_F=0.15$, which we note is consistent with the results obtained by PINNs and also agrees with the critical temperature observed in the experiment \cite{Ku}. Due to the large fluctuations in determining the critical temperature in the bosonic sector, this method of inferring the critical temperature has large fluctuations. However, this method can still provide auxiliary evidence for the existence of phase transition in the Fermi system. The red dot in the figure is the critical temperature given for $N=20$ based on contact interaction in the article \cite{Gilbreth}.

In Fig. \ref{3535specific}, we show the specific heat per particle for the Fermi system with $N_\uparrow=N_\downarrow=35$. The blue dots are the specific heat per particle obtained after PINNs training, where we included the ground state energy given by the constant energy semi-extrapolation in the PINN training. The black line is the specific heat per particle of the ideal Fermi gas. The red arrow in the figure is the critical temperature observed by Ku et al., in the experiment \cite{Ku}. It is worth noting that although the specific heat curve in the figure looks very good, considering the fluctuations of the specific heat, it will still lead to obvious error bars when determining the critical temperature.

It is worth noting that simulating the thermodynamic properties of ultracold Fermi gases in the unitary limit in a harmonic trap is not an easy task due to the difficulty caused by the fermion sign problem. In the article \cite{Gilbreth}, in order to suppress the fermion sign problem \cite{Bulgac}, the contact interaction was used, and the auxiliary-field Monte Carlo (AFMC) method was applied to simulate the specific heat at low temperatures for the spin-balanced case of $N_\uparrow=N_\downarrow=10$. For the AFMC method, the computational complexity involved in increasing the number of fermions is much larger than that involved in the fictitious identical particle thermodynamics. Therefore, only the low-temperature results for $N_\uparrow=N_\downarrow=10$ were given in the article \cite{Gilbreth}. In addition, there are also greater difficulties for higher temperatures, while in the fictitious identical particle method, higher temperatures do not bring additional difficulties. Due to the high efficiency and reliability of the fictitious identical particle combined with PINNs to simulate the thermodynamic properties of the Fermi system, we have systematically simulated the phase transitions of different number of fermions compared with the article \cite{Gilbreth}. In the article \cite{latticeMC}, the contact interaction was also used, and the lattice Monte Carlo was used to simulate the ground state energy of up to 70 fermions. This method faces great difficulties for finite temperature, and it is also difficult to deal with the more realistic short-range interactions considered here. In lattice Monte Carlo, space needs to be discretized; for the short-range interactions considered here, this discretization faces great challenges. Finally, it is worth pointing out that in methods such as AFMC, under the contact interaction, the fermion sign problem can be partially suppressed when the number of fermions with different spins is the same, but when the number of fermions with different spins is different, the fermion sign problem is still very serious, and this situation is especially important in the simulation of phenomena such as magnetism. In our method, there is no additional difficulty caused by the different number of fermions with different spins, so it is expected to be applied in a wider range of situations \cite{Schunck,Shin}.

The research of this section fully demonstrates that the combination of fictitious identical particles and PINNs provides a valuable tool for simulating the phase transitions of Fermi systems from first principles. Due to the essential overcoming of the fermion sign problem from the perspective of computational complexity, this method can play a unique role especially in simulating large-scale Fermi systems. As a comparison, in our simulation of the Fermi system with periodic potential, no phase transition was observed. The reason is that we consider the repulsive interaction, and it is not surprising that no phase transition of the Fermi condensate is simulated under the selected parameter conditions. We believe that the method developed in this article is promising for future simulations of the phase transitions of Fermi gases in the unitary limit with periodic potential.

\section{Summary and discussion}
\label{summary}

In summary, in this work, we combine fictitious identical particles with PINNs to simulate the thermodynamic properties and phase transitions of Fermi systems based on the thermodynamic properties of the bosonic sector and the physical information  
$\left.\frac{\partial E(\xi,T)}{\partial T}\right|_{T=0}=0$. Our simulations of three different typical Fermi systems fully demonstrate the efficiency and reliability of this combined method. In the current demonstration, whether it is the continuous Fermi-Hubbard model or the ultracold Fermi gas in the unitary limit with short-range interaction in the harmonic trap, on a small server, we are able to exceed the scale of the Fermi system that can be simulated with much more computational resources in the number of simulated fermions, compared to other methods.

In this work, we generalize the new recursion formula for identical bosons by Feldman and Hirshberg \cite{HirshImprove} to the PIMD \cite{Xiong-quadratic} of fictitious identical particle thermodynamics. In the PIMD simulation of the bosonic sector, the computational complexity is $O(N^2+PN)$, where $N$ is the number of identical particles and $P$ is the number of beads for each quantum particle when the partition function of the quantum system is transformed into path integral form. Since we need to perform PIMD simulations for different $\xi$ and different temperatures in the bosonic sector, the total computational complexity is $O(M(N^2+PN))$ for analyzing the thermodynamic properties of the Fermi system. Here $M$ is the number of independent simulations required for different $\xi$ and $T$ in the bosonic sector. In this paper, $M$ is usually between 250 and 300. This is the price to pay for accurately simulating the thermodynamic properties of the Fermi system. However, this price is worth it, because once we have accumulated enough data in the bosonic sector, we can obtain the thermodynamic properties of the Fermi system from zero temperature to finite temperature continuously through PINNs. In comparison, other methods often need to be carried out independently for different temperatures of the Fermi system. This advantage has been proved to be valuable for simulating specific heat and phase transition in this work. The large amount of data accumulated in the bosonic sector also helps to greatly suppress the energy fluctuations of the Fermi system simulation. Since all $M$ data are used, the energy fluctuations will be suppressed at the $1/\sqrt M$  level. For this reason, no error bars caused by fluctuations are given in the energy results in this paper. Of course, the combination of fictitious identical particles and PINNs can still lead to systematic errors. This systematic error is difficult to evaluate due to the lack of exact energy for comparison. In this case, the comparison of PINNs, constant energy semi-extrapolation and isothermal extrapolation helps to evaluate whether the simulation results are reasonable.

In this work, we only analyze the energy and specific heat of the Fermi system. In fact, for any physical quantity such as the density distribution $n(\xi,T,{\textbf r})$, we can prove that  
$\left.\frac{\partial n(\xi,T,{\textbf r})}{\partial T}\right|_{T=0}=0$. This exact relationship can also be used for PINNs. In future work, we will apply the method of this paper to important physical quantities such as density distribution, momentum distribution, density correlation function, static structure factor and static linear density response function. Compared with energy, these physical quantities also contain continuous variables, and we expect PINNs to be particularly valuable in the simulation of these physical quantities. For neural networks, accumulating more data during training helps the accuracy of extrapolation. In addition, simple feedforward neural networks are used in this work. It is also a topic worth studying to apply various powerful methods developed in deep learning to physics-informed neural networks to improve the accuracy of the simulation.

The combination of fictitious identical particles and PINNs provides a common platform for various theoretical simulation results and experimental observations, because imposing physical information on PINNs by various means helps us to continuously deepen our precise understanding of Fermi systems. The main purpose of this paper is to establish a new method for simulating Fermi systems, rather than to systematically study a particular quantum system. As Dornheim pointed out in the article \cite{Dornheim1,Dornheim2}, once the $\xi$ extrapolation method is further developed, it will have a wide range of application prospects in the study of the thermodynamic properties of quantum systems, such as strongly coupled electron liquid, fractional quantum Hall effect, ultracold fermionic gases and brown dwarfs, etc.. We believe that the combination of fictitious identical particles and PINNs is laying the foundation for these broad and important application prospects.

\begin{acknowledgments}
This work is partly supported by the National Natural Science Foundation of China under grant numbers 11175246, and 11334001. This work has received funding from Hubei Polytechnic University.
\end{acknowledgments}

\textbf{DATA AVAILABILITY}

The data that support the findings of this study are available from the corresponding author upon reasonable request. The code is available on GitHub at \\
https://github.com/xiongyunuo/MD-Lab \\
https://github.com/xiongyunuo/PINNFermion.


\begin{thebibliography}{10}


\bibitem{ceperley} D. M. Ceperley, “Path Integral Monte Carlo Methods for Fermions”, in \textit{Monte Carlo and Molecular Dynamics of Condensed Matter Systems}, edited by K. Binder and G. Ciccotti (Editrice Compositori, Bologna, Italy, 1996).

\bibitem{troyer} M.~Troyer and U. J.~Wiese, 
“Computational Complexity and Fundamental Limitations to Fermionic Quantum Monte Carlo Simulations,” 
{\text{Phys. Rev. Lett.} \textbf{94}, 170201} (2005).

\bibitem{WDM} T. Dornheim, S. Groth, and M. Bonitz, “The uniform electron gas at warm dense matter conditions,” Phys. Rep. \textbf{744}, 1 (2018).

\bibitem{Dornheim-PRE} T.~Dornheim, 
“The Fermion sign problem in path integral Monte Carlo simulations: quantum dots, ultracold atoms, and warm dense matter,” 
\text{Phys. Rev. E}~\textbf{100}, 023307 (2019).

\bibitem{Alex} A. Alexandru, G. Basar, P. F. Bedaque, and N. C. Warrington, 
“Complex paths around the sign problem,” 
Rev. Mod. Phys. \textbf{94}, 015006 (2022).

\bibitem{nodes} D. M. Ceperley, “Fermion nodes,” J. Stat. Phys. \textbf{63}, 1237 (1991).

\bibitem{Helium}  D. M. Ceperley, “Path-integral calculations of normal liquid 
$^3$He,” Phys. Rev. Lett. \textbf{69}, 331 (1992).

\bibitem{Militzer} B. Militzer, E. L. Pollock, and D. M. Ceperley, “Path integral Monte Carlo calculation of the momentum distribution of the homogeneous electron gas at finite temperature,” High Energy Dens. Phys. \textbf{30}, 13 (2019).

\bibitem{Mak} C. H. Mak, R. Egger, and H. Weber-Gottschick, “Multilevel blocking approach to the fermion sign problem in path-integral Monte Carlo simulations,” Phys. Rev. Lett. \textbf{81}, 4533 (1998).

 
\bibitem{Rubtsov} A. N. Rubtsov, V. V. Savkin, A. I. Lichtenstein, “Continuous-time quantum Monte Carlo method for fermions,” Phys. Rev. B \textbf{72}, 035122 (2005).
 
 \bibitem{diMonte} N. V. Prokof’ev and B. V. Svistunov, “Bold diagrammatic Monte Carlo technique: When the sign problem is welcome,” Phys. Rev. Lett. \textbf{99}, 250201 (2007).

 
 \bibitem{Burov2} E. Burovski, N. V. Prokof’ev, B. V. Svistunov, and M. Troyer, “Critical Temperature Curve in BEC-BCS Crossover,” \text{Phys. Rev. Lett.}~  \textbf{101}, 090402 (2008).
 
 \bibitem{diMonte2} P. C. Hou, B. Z. Wang, K. Haule, Y. Deng, and K. Chen, “Exchange-correlation effect in the charge response of a warm dense electron gas,” Phys. Rev. B \textbf{106}, L081126 (2022).


\bibitem{Blunt} N. S. Blunt, T. W. Rogers, J. S.  Spencer, and W. M. Foulkes,
“Density-matrix quantum Monte Carlo method,” Phys.
Rev. B \textbf{89}, 245124 (2014). 

\bibitem{Malone} F. D. Malone, N. S. Blunt, James J. Shepherd, D. K. K. Lee,  J. S. Spencer, and  W. M. C. Foulkes, “Interaction Picture Density Matrix Quantum Monte Carlo,” J. Chem. Phys. \textbf{143}, 044116 (2015).

\bibitem{Schoof1} T. Schoof, M. Bonitz, A. V. Filinov, D. Hochstuhl and
J. W. Dufty, “Configuration Path Integral Monte Carlo,”
Contrib. Plasma Phys. \textbf{51}, 687 (2011).

\bibitem{Schoof2} T. Schoof, S. Groth,  and M. Bonitz, “Towards \textit{ab initio} thermodynamics of the electron gas at strong degeneracy,” Contrib. Plasma Phys. \textbf{55}, 136 (2015).

\bibitem{Schoof3} T. Schoof, S. Groth, J. Vorberger,  and M. Bonitz, “\textit{Ab Initio} Thermodynamic Results for the Degenerate Electron Gas at Finite Temperature,” Phys. Rev. Lett. \textbf{115}, 130402 (2015).

\bibitem{Yilmaz} A. Yilmaz, K. Hunger,  T. Dornheim, S. Groth, and  M. Bonitz, “Restricted configuration path integral Monte Carlo,” J. Chem. Phys. \textbf{153}, 124114 (2020).

\bibitem{PB1} T. Dornheim, S. Groth, A. Filinov, and M. Bonitz, “Permutation blocking path integral Monte Carlo: a highly efficient approach to the simulation of strongly degenerate non-ideal fermions,” New J. Phys. \textbf{17}, 073017 (2015).

\bibitem{PB2} T. Dornheim, T. Schoof, S. Groth, A. Filinov, and M. Bonitz, “Permutation blocking path integral Monte Carlo approach to the uniform electron gas at finite temperature,” J. Chem. Phys. \textbf{143},  204101 (2015).

\bibitem{Groth} S. Groth, T. Dornheim, T. Sjostrom, F. D. Malone, W. M. C. Foulkes, and M. Bonitz, “\textit{Ab initio} exchange-correlation free energy of the uniform electron gas at warm dense matter conditions,” Phys. Rev. Lett. \textbf{119}, 135001 (2017).


\bibitem{Joonho} J. Lee, M. A. Morales, and F. D. Malone, “A phaseless auxiliary-field quantum Monte Carlo perspective on the uniform electron gas at finite temperatures: Issues, observations, and benchmark study,” J. Chem. Phys. \textbf{154}, 064109 (2021).

\bibitem{SWZhang} J. Carlson, S. Gandolfi, K. E. Schmidt, S.  Zhang, “Auxiliary-field quantum Monte Carlo method for strongly paired fermions,” Phys. Rev. A \textbf{84}, 061602 (2011).

\bibitem{QinMP} M. Qin, H. Shi,  and S. Zhang,  “Benchmark study of the two-dimensional Hubbard model with auxiliary-field quantum Monte Carlo method,” Phys.  Rev. B \textbf{94}, 085103 (2016).

\bibitem{DornheimMod} T. Dornheim, M. Invernizzi, J. Vorberger, and B. Hirshber, 
“Attenuating the fermion sign problem in path integral Monte Carlo simulations using the Bogoliubov inequality and thermodynamic integration,” 
J. Chem. Phys. \textbf{153}, 234104 (2020).

\bibitem{XiongFSP} Y. Xiong and H. Xiong, “On the thermodynamic properties of fictitious identical particles and the application to fermion sign problem,” J. Chem. Phys. \textbf{157}, 094112 (2022).  

\bibitem{Xiong-xi} Y. Xiong and H. Xiong, “On the thermodynamics of fermions at any temperature based on parametrized partition function,” Phys. Rev. E  \textbf{107}, 055308 (2023).

\bibitem{Dornheim1} T. Dornheim, P. Tolias, S. Groth, Z. A. Moldabekov, J. Vorberger, and B. Hirshberg, “Fermionic physics from \textit{ab initio} path integral Monte Carlo simulations of fictitious identical particles,”  J. Chem. Phys. \textbf{159}, 164113 (2023).

\bibitem{Dornheim2} T. Dornheim, S. Schwalbe, Z. A.  Moldabekov, J. Vorberger, and P. Tolias, “\textit{Ab initio} path integral Monte Carlo simulations of the uniform electron gas on large length scales,” J. Phys. Chem. Lett. \textbf{15}, 1305 (2024).


\bibitem{HirshPIMD} B. Hirshberg, V. Rizzi, and M. Parrinello, “Path integral molecular dynamics for bosons,” Proc. Natl. Acad. Sci. USA \textbf{116}, 21445 (2019).

\bibitem{HirshImprove} Y. M. Y. Feldman and B. Hirshberg, “Quadratic Scaling Bosonic Path Integral Molecular Dynamics,” J. Chem. Phys. \textbf{159}, 154107 (2023).

 \bibitem{Deuterium}   C. W. Myung, B. Hirshberg, and M. Parrinello, “Prediction of a supersolid phase in high-pressure deuterium,” \text{Phys. Rev. Lett.} \textbf{128}, 045301 (2022).

\bibitem{Xiong-spinor} Y. Yu, S. Liu, H. Xiong, and Y. Xiong, “Path integral molecular dynamics for thermodynamics and Green's function of ultracold spinor bosons,” J. Chem. Phys. \textbf{157}, 064110 (2022).

\bibitem{Xiong-Momentum} Y. Xiong and  H. Xiong, 
“Numerical calculation of Green's function and momentum distribution for spin-polarized fermions by path integral molecular dynamics,” 
J. Chem. Phys. \textbf{156}, 204117 (2022).

\bibitem{Xiong-magnetic} Y. Xiong and H. Xiong, “Path integral and winding number in singular magnetic field,” Eur. Phys. J. Plus \textbf{137}, 550 (2022).

\bibitem{Xiong-anyon} Y. Xiong and H. Xiong, “Path integral molecular dynamics for anyons, bosons, and fermions,” Phys. Rev. E \textbf{106}, 025309 (2022).

\bibitem{Xiong-Green} Y. Xiong and H. Xiong, “Path integral molecular dynamics simulations for Green’s function in a system of identical bosons,” J. Chem. Phys. \textbf{156}, 134112 (2022).

\bibitem{CeperleyBook} R. M. Martin, L. Reining, and D. M. Ceperley, \textit{Interacting Electrons: Theory and Computational Approaches} (Cambridge University Press, Cambridge, UK, 2016).

\bibitem{Feynman} R. P. Feynman and R. H.  Albert,  \textit{Quantum mechanics and path integrals}  (McGraw-Hill, New York, 1965).

\bibitem{Tuckerman} M. E.~Tuckerman, \textit{Statistical mechanics: theory and molecular simulation} (Oxford University, New York, 2010).

\bibitem{barker} J. A.~Barker, “A quantum-statistical Monte Carlo method; path integrals with boundary conditions,” J. Chem. Phys. \textbf{70}, 2914 (1979).

\bibitem{Morita} T. Morita, “Solution of the Bloch Equation for Many-Particle Systems in Terms of the Path Integral,” J. Phys. Soc. Japan. \textbf{35}, 980 (1973).

\bibitem{CeperleyRMP} D. M. Ceperley, “Path integrals in the theory of condensed helium,” Rev. Mod. Phys. \textbf{67}, 279 (1995).

\bibitem{Burov1} M. Boninsegni, N. V. Prokof’ev, and B. V. Svistunov, “Worm Algorithm for Continuous-Space Path Integral Monte Carlo Simulations,” \text{Phys. Rev. Lett.} ~\textbf{96}, 070601 (2006). 

\bibitem{Burov1b} M. Boninsegni, N. V. Prokof’ev, and B. V. Svistunov, “Worm algorithm and diagrammatic Monte Carlo: A new approach to continuous-space path integral Monte Carlo simulations,” Phys. Rev. E \textbf{74}, 036701 (2006).


\bibitem{neuralbook} Y. LeCun, Y. Bengio, and G. Hinton, “Deep learning,” Nature \textbf{521}, 436 (2015).


\bibitem{PINN1} G. E. Karniadakis, I. G. Kevrekidis, L. Lu,  P. Perdikaris, S. Wang, L.  Yang, “Physics-informed machine learning,” Nat. Rev. Phys. \textbf{3}, 422 (2021).

\bibitem{PINN2} S. Cuomo, V. S. Di Cola, F. Giampaolo, G. Rozza, M. Raissi, F.  Piccialli, “Scientific machine learning through physics–informed neural networks: Where we are and what’s next,” J. Sci. Comput. \textbf{92}, 88 (2022).

\bibitem{PINN3} M. Raissi, P. Perdikaris, G. E. Karniadakis, “Physics-informed neural networks: A deep learning framework for solving forward and inverse problems involving nonlinear partial differential equations,” J. Comput. Phys. \textbf{378}, 686 (2019).

\bibitem{lattice} I. Bloch, “Ultracold quantum gases in optical lattices,” Nat. Phys. \textbf{1}, 23 (2005).

\bibitem{Bloch} I. Bloch, J. Dalibard, and W. Zwerger, “Many-body physics with ultracold gases,” Rev. Mod. Phys. \textbf{80}, 885 (2008).

\bibitem{Fermilattice} T. Esslinger, “Fermi-Hubbard physics with atoms in an optical lattice,” Annu. Rev. Condens. Matter Phys. \textbf{1}, 129 (2010).


\bibitem{LeBlanc} J. P. F. LeBlanc, et al., “Solutions of the Two-Dimensional Hubbard Model: Benchmarks and Results from a Wide Range of Numerical Algorithms,” Phys. Rev. X \textbf{5}, 041041 (2015).

\bibitem{Qin} M. Qin, T. Schäfer, S. Andergassen, P. Corboz, and E. Gull, “The Hubbard model: A computational perspective,” Annu. Rev. Condens. Matter Phys. \textbf{13}, 275 (2022).


\bibitem{Xiong-quadratic} Y. Xiong, S. Liu, and H. Xiong, “Quadratic scaling path integral molecular dynamics for fictitious identical particles and its application to fermion systems,” preprint arXiv:2401.00274 (2023).


\bibitem{xiong-arxiv} Y. Xiong and H. Xiong, “Parametrized path integral formulation for large fermion systems,”  preprint arXiv:2208.13777 (2022).


\bibitem{Bertsch} T. Papenbrock, G. F. Bertsch, “Pairing in low-density Fermi gases,” Phys. Rev. C \textbf{59}, 2052 (1999).

\bibitem{Ho} T. L. Ho, “Universal thermodynamics of degenerate quantum gases in the unitarity limit,” Phys. Rev. Lett. \textbf{92}, 090402 (2004).


\bibitem{Strinati} G. C. Strinati, P. Pieri, G. Röpke, P. Schuck, and M. Urban, “The BCS–BEC crossover: From ultra-cold Fermi gases to nuclear systems,” Phys. Rep. \textbf{738}, 1 (2018).



\bibitem{Thomas} J. E. Thomas, J. Kinast,  and A. Turlapov, “Virial theorem and universality in a unitary Fermi gas,” Phys. Rev. Lett. \textbf{95}, 120402 (2005).

\bibitem{Sagi} Y. Sagi, T. E. Drake, R. Paudel, and D. S. Jin, “Measurement of the homogeneous contact of a unitary Fermi gas,” Phys. Rev. Lett. \textbf{109}, 220402 (2012).

\bibitem{Ku} M. J. Ku, A. T. Sommer, L. W. Cheuk, and M. W. Zwierlein, “Revealing the superfluid lambda transition in the universal thermodynamics of a unitary Fermi gas,” Science \textbf{335}, 563 (2012).

\bibitem{Horikoshi} M. Horikoshi, S. Nakajima, M. Ueda, and T. Mukaiyama,  “Measurement of universal thermodynamic functions for a unitary Fermi gas,” Science \textbf{327}, 442 (2010).

\bibitem{LiX} X. Li, S. Wang, X. Luo, Y. Y. Zhou, K. Xie, H. C. Shen, ...  J. W. Pan, “Observation and quantification of the pseudogap in unitary Fermi gases,” Nature \textbf{626}, 288 (2024).


\bibitem{Gilbreth} C. N. Gilbreth, and Y. Alhassid, “Pair condensation in a finite trapped Fermi gas,”  Phys. Rev. A  \textbf{88},  063643 (2013).


\bibitem{latticeMC} M. G. Endres, D. B. Kaplan, J. W. Lee,  and A. N. Nicholson, “Lattice Monte Carlo calculations for unitary fermions in a harmonic trap,” Phys. Rev. A \textbf{84}, 043644 (2011).


\bibitem{Mukherjee} A. Mukherjee, and Y. Alhassid, “Configuration-interaction Monte Carlo method and its application to the trapped unitary Fermi gas,” Phys. Rev. A \textbf{88},  053622 (2013).  

\bibitem{Carlson} J. Carlson and S. Gandolfi, “Predicting energies of small clusters from the inhomogeneous unitary Fermi gas,” Phys. Rev. A \textbf{90},  011601 (2014). 

\bibitem{FNDMC} D. Blume, J. von Stecher, and C. H. Greene, “Universal properties of a trapped two-component Fermi gas at unitarity,” Phys. Rev. Lett. \textbf{99}, 233201 (2007).

\bibitem{GFMC} S. Y. Chang and G. F. Bertsch, “Unitary Fermi gas in a harmonic trap,” Phys. Rev. A \textbf{76}, 021603 (2007).


\bibitem{GaussianInt} P. Jeszenszki, A. Y. Cherny, and J. Brand, “The s-wave scattering length of a Gaussian potential,” Phys. Rev. A \textbf{97}, 042708 (2018).

\bibitem{Bulgac} A. Bulgac, J. E. Drut, and P. Magierski, “Spin 1/2 Fermions in the Unitary Regime: A Superfluid of a New Type,” Phys. Rev. Lett. \textbf{96}, 090404 (2006).


\bibitem{Schunck} C. H. Schunck, Y. Shin, A. Schirotzek, M. W. Zwierlein, W.  Ketterle,  “Pairing without superfluidity: The ground state of an imbalanced Fermi mixture,” Science \textbf{316}, 867 (2007).


\bibitem{Shin} Y. I. Shin, C. H. Schunck, A. Schirotzek, W. Ketterle, “Phase diagram of a two-component Fermi gas with resonant interactions,” Nature \textbf{451}, 689 (2008).

\end{thebibliography}
\end{document}